\documentstyle[jkas2,epsfig]{article}

\newcommand{\hmpc}{\;h^{-1}{\rm Mpc}}
\newcommand{\hmpccc}{\;h^{3}{\rm Mpc}^{-3}}
\newcommand{\fxcgs}{\rm \, \times 10^{-12} \;erg\;cm^{-2}\;s^{-1}}
\newcommand{\lxcgs}{{\rm \, \times 10^{44}\;} h^{-2} {\rm erg\;s^{-1}}}
\newcommand{\lx}{L_{\rm x}}
\newcommand{\fx}{f_{\rm x}}

\runningauthor{LEE AND PARK}
\runningtitle{CORRELATION FUNCTIONS OF CLUSTERS OF GALAXIES}
\beginpage{1}
\endpage{10}

\begin{document}

\title{CORRELATION FUNCTIONS OF THE ABELL, APM, AND X-RAY CLUSTERS
OF GALAXIES}

\author{SUNGHO LEE AND CHANGBOM PARK}

\address{Astronomy Program in SEES, Seoul National University, Seoul, 151-742 Korea \\
{\it E-mail: leesh@astro.snu.ac.kr and cbp@astro.snu.ac.kr}}

\address{\normalsize{\it (Received ???. ??, 2002; Accepted ???. ??,2002)}}

\abstract{
We have measured the correlation functions of the
optically selected clusters of galaxies in the Abell and the APM catalogs,
and of the X-ray clusters in
the X-ray-Brightest Abell-type Clusters of galaxies (XBACs) catalog and
the Brightest Clusters Sample (BCS).
The same analysis method and the same method of characterizing
the resulting correlation functions are applied to all observational samples.
We have found that the amplitude of the correlation function
of the APM clusters is much higher than what has been previously claimed,
in particular for richer subsamples.
The correlation length of the APM clusters with the richness
${\cal R} \ge 70$ (as defined by the APM team) is
found to be $r_0 = 25.4^{+3.1}_{-3.0} \hmpc$.
The amplitude of correlation function is about
2.4 times higher than that of Croft et al. (1997).
The correlation lengths of the Abell clusters
with the richness class $RC \ge 0$ and 1 are measured to be
$r_0 = 17.4^{+1.2}_{-1.1}$ and $21.0^{+2.8}_{-2.8} \hmpc$,
respectively, which is consistent with our results for the APM sample
at the similar level of richness.
The richness dependence of cluster correlations is found to be
$r_{0} =0.40 d_{c} + 3.2$ where $d_c$ is the mean intercluster separation.
This is identical in slope with the Bahcall \& West (1992)'s estimate,
but is inconsistent with the weak dependence of Croft et al. (1997).
The X-ray bright Abell clusters in the XBACs catalog and
the X-ray selected clusters in the BCS catalog show strong clustering.
The correlation length of the
XBACs clusters with $\lx \ge 0.65 \lxcgs$ is $30.3^{+8.2}_{-6.5} \hmpc$,
and that of
the BCS clusters with $\lx \ge 0.70 \lxcgs$ is $30.2^{+9.8}_{-8.9} \hmpc$.
The clustering strength of the X-ray clusters is much weaker than
what is expected from the optical clusters.}

\maketitle


\section{INTRODUCTION} \label{intro}

Rich clusters of galaxies are used to study the large-scale
structure of the universe on scales larger than about $10 \hmpc$
($h = {\rm H_0\;/\;100\;km\;s^{-1}\;Mpc^{-1}}$).
To measure the clustering strength of clusters
the spatial two-point correlation function (CF) of
the Abell clusters (\cite{abe58}; Abell, Corwin, \& Olowin 1989)
has been estimated by many authors
(Bahcall \& Soneira 1983;
\cite{huc90}; Postman, Huchra, \& Geller 1992; \cite{pea92}; \cite{mil99};
see \cite{bah88} for a review),
and has been found to be consistent with the power law
\begin{equation}
\xi_{cc}(r) = \left( {r \over r_0} \right)^{-\gamma}
\end{equation}
with the correlation length $r_0 \simeq 20 \sim 26 \hmpc$
and with the index $\gamma \simeq 2$.
Recently new catalogs of clusters have been obtained by
automated selection from the Edinburgh-Durham Southern Galaxy Catalog
and the Automatic Plate Measuring (APM) Galaxy Survey
(\cite{lum92}; \cite{dal97}).
Redshift data of these clusters
(\cite{col95}; \cite{dal94b}) have been used to
estimate the CF (\cite{nic92}; \cite{dal92}; \cite{dal94a}).
The correlation lengths $r_0$ measured for these new cluster samples have
been reported to be $14 \lesssim r_0 \lesssim 16 \hmpc$, which is
much smaller than that of the Abell clusters.
There are also more recent new catalogs such as
Northern Sky Optical Cluster Survey based on Digitized Second Palomar Sky Survey (DPOSS)
(\cite{gal00}) and
Sloan Digital Sky Survey Cut-and-Enhance (SDSS-CE) galaxy cluster catalog (\cite{got02}).
They need, however, either more follow-up confirmations or spectroscopic observations
before being able to be used to study the spatial clustering.

It has been argued that the counting radius $r_c = 1.5 \hmpc$
for the Abell clusters is so large that the catalog contains
serious projection effects, which cause artificial
line-of-sight correlations (\cite{sut88}; \cite{efs92}).
It has also been pointed out that
the intrinsically subjective nature of the Abell catalog can cause
problems in homogeneity and statistical completeness (\cite{nic92}).
However, Bahcall \& West (1992) have claimed that the
discrepancy between the Abell and the APM cluster CFs can be explained by
the richness dependence of cluster correlation amplitudes $r_0 = 0.4 d_c$,
where $d_c = n_c ^{-1/3}$ is the mean intercluster separation and
$n_c$ is the mean space density of clusters.
But Croft et al. (1997) have analyzed richness subsamples
of the APM clusters and argued that there is only
a weak dependence of correlation amplitudes on the cluster richness.

\begin{figure*}[t]
\plotfiddle{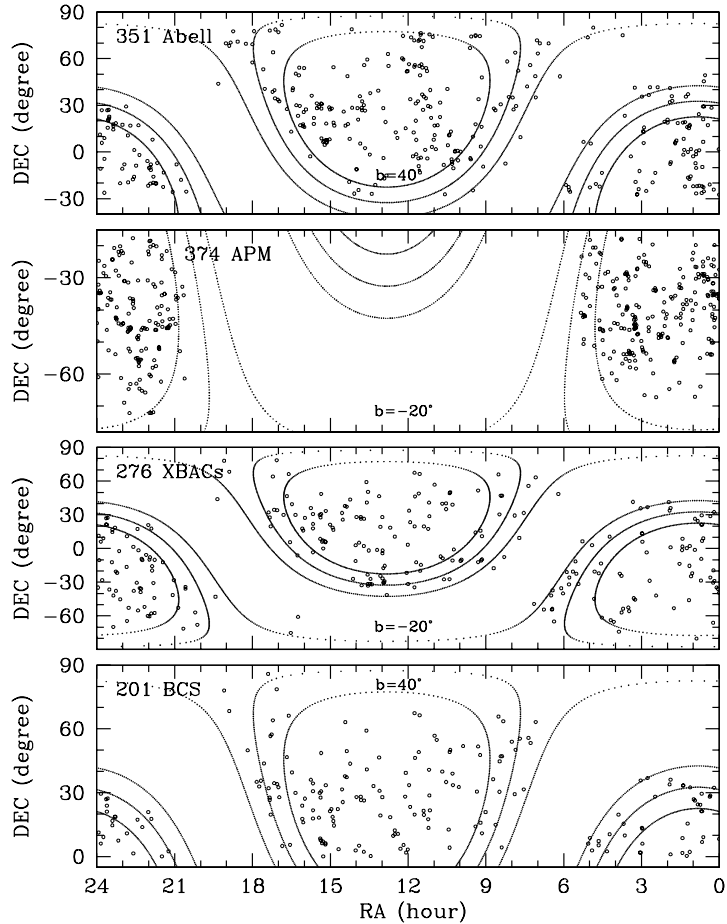}{12.5cm}{0}{67}{67}{-210}{-130}
\caption{\footnotesize
Distribution of clusters of galaxies on the sky
with measured redshifts;
351 Abell clusters, 374 APM clusters, 276 XBACs clusters, and 201 BCS clusters.
The dotted curves represent the galactic latitudes
$b = \pm40\arcdeg$, $\pm30\arcdeg$, and $\pm20\arcdeg$.
\label{fig_sky}}
\end{figure*}

It is, therefore, necessary to understand why there have been
disagreements in the cluster correlation length $r_0$ and on the relation
between $r_0$ and $d_c$.
In the previous studies individual observational samples have been analyzed
by different authors who used different methods of analysis.
In the current situation where the cluster clustering results
from different studies do not agree,
it is important to reanalyze and compare all preciously studied
samples in an impartial way.
In this paper we apply methods of calculating  and
characterizing the CFs to all cluster samples
in a same way to remove any relative biases.
It is hoped that in this objective and consistent way of analysis
we could find the common characteristics of, and
intrinsic differences in the spatial distributions
of various cluster samples.

On the other hand, rich clusters of galaxies often
have diffuse intracluster gas trapped in their potential wells.
The thermal X-ray flux from the intracluster gas, which is
heated to temperature
of a few $10^7\;{\rm K}$, is proportional to the square of the ion density,
and thus is more confined to the center of the clusters than the
projected galaxy distribution.
Therefore, the X-ray selected clusters  are expected to have negligible
projection effects (\cite{bri93}; \cite{ebe96}).
Ebeling et al. (1996) have cross-correlated the clusters in the ACO catalog
(\cite{abe89}) with the {\it ROSAT} X-ray sources (\cite{tru93})
and created the X-ray-Brightest Abell-type Clusters of galaxies (XBACs) catalog.
More recently, Ebeling et al. (1998) have published the
Brightest Clusters Sample (BCS) which is a true X-ray selected sample
from the {\it ROSAT} All-Sky Survey.
It should be interesting to compare the clustering strengths of these
X-ray cluster samples with those of optically selected samples to further extend
our knowledge on the clustering properties of clusters of galaxies.

\begin{deluxetable}{ccccccccc}
\tablecolumns{5}
\footnotesize
\tablecaption{
Richness subsamples of the Abell and the APM clusters
\label{table_op}}
\tablewidth{0pt}
\tablehead{
\colhead{} & \colhead{} &
\colhead{$n_c$} & \colhead{$r_0$} & \colhead{} \nl
\colhead{Richness} & \colhead{$N_c$} & \colhead{($\!\hmpccc$)} &
\colhead{($\!\hmpc$)} & \colhead{$\gamma$}
}
\startdata
\tablevspace{0.9mm}
\multicolumn{5}{c}{\small Abell} \nl
\tableline
\tablevspace{1mm}
 $RC \ge 0$ & 232 & $1.74 \times 10^{-5}$
& $17.4^{+1.2}_{-1.1}$ & $1.89^{+0.15}_{-0.16}$ \nl
\noalign{\smallskip}
 $RC \ge 1$ & 110 & $0.68 \times 10^{-5}$
& $21.0^{+2.8}_{-2.8}$ & $1.91^{+0.34}_{-0.36}$ \nl
\tablevspace{0.7mm}
\tableline
\tablevspace{2mm}
\multicolumn{5}{c}{\small APM} \nl
\tableline
\tablevspace{1mm}
 ${\cal R} \ge 50$ & 218 & $2.62 \times 10^{-5}$
& $15.4^{+0.8}_{-0.8}$ & $2.49^{+0.17}_{-0.16}$ \nl
\noalign{\smallskip}
 ${\cal R} \ge 60$ & 139 & $1.66 \times 10^{-5}$
& $22.3^{+1.6}_{-1.6}$ & $1.96^{+0.16}_{-0.17}$ \nl
\noalign{\smallskip}
 ${\cal R} \ge 70$ &  73 &  $0.93 \times 10^{-5}$
& $25.4^{+3.1}_{-3.0}$ & $2.01^{+0.29}_{-0.30}$ \nl
\noalign{\smallskip}
 ${\cal R} \ge 80$ &  42 &  $0.57 \times 10^{-5}$
& $26.8^{+7.0}_{-5.7}$ & $2.03^{+0.58}_{-0.67}$ \nl
\enddata
\end{deluxetable}


\section{DATA} \label{data}


\subsection{The Abell Clusters} \label{abelldata}

Postman et al. (1992) have published a complete redshift catalog of
351 Abell clusters with
richness class $RC \ge 0$, declination $\delta \ge -27\fdg5$,
and tenth-ranked galaxy magnitude $m_{10} \le 16.5$.
We restrict our full sample to 233 clusters with $|b| \ge 40\arcdeg$
to avoid serious incompleteness at low galactic latitude (\cite{efs92}).
In this region with solid angle $\Omega = 3.48$ sr,
clusters are nearly uniformly distributed over Galactic latitude.
Two richness subsamples with $RC \ge 0$ and $RC \ge 1$ are adopted.

Assuming the Einstein-de Sitter universe,
we transform redshifts into comoving distances by using the relation

\begin{equation}
d(z) = 6000 ( { 1 - 1 / \sqrt{1+z}  } ) \hmpc \; .
\end{equation}
Since we use the same $d(z)$ relation to all cluster catalogs,
the resulting CFs are not biased with respect to
one another due to the choice of cosmology.
Furthermore, most other authors have also adopted
the standard Einstein-de Sitter model in their calculation.

To estimate the selection function of the survey in the radial direction,
we first compute the absolute magnitude $M_{10}$ corresponding to
$m_{10}$ of each cluster at a redshift $z$  from the relation
\begin{equation}
m - M = 5 \log \, [ \, d(z)(1+z) \, ] +25 + K(z) \; .
\end{equation}
The K-correction $K(z)$ is $1.122z$ for the red magnitudes
of the northern Abell clusters
and $4.14z - 0.44z^2$ for the visual magnitudes
of the southern ACO clusters (\cite{ebe96}).
Then the maximum distance $D_{max,i}$ to which each cluster
could be observed for the survey magnitude limit
$m_{10,lim} = 16.5$ can be calculated.
And the selection function is given by
\begin{equation}
S(d) = {3\over \Omega} \sum_{i; \atop {D_{max,i} \ge d}}
{ 1 \over {D^{3}_{max,i}}} \; .
\end{equation}
Using the distribution of clusters smoothed in the redshift space
as the radial selection function does not make much difference in our results.
We limit the sample to the distance range
from $d_{in} = 50 \hmpc$ to $d_{out} = 500 \hmpc$,
and the resulting sample includes 232 and 110 clusters for
$RC \ge 0$ and $RC \ge 1$, respectively (see Table~\ref{table_op}).
The region closer than $d_{in}=50 \hmpc$
is removed to improve the completeness of the Abell
cluster sample and to avoid the possible local effects.
Large variation of the locations of the inner and outer boundaries makes
negligible effect on our results.

We estimate the comoving space density of clusters in each sample
using the method described in Section 2.3 of Efstathiou et al. (1992).
The mean space density of the Abell cluster sample with
$RC \ge 0$ is $1.74 \times 10^{-5} \hmpccc$ while
Postman et al. (1992)'s estimate is $1.2 \times 10^{-5} \hmpccc$ and
Efstathiou et al. (1992)'s is $1.9 \times 10^{-5}\hmpccc$.
The density of our $RC \ge 1$ subsample,
$n_c = 6.79 \times 10^{-6}\hmpccc$, agrees well with Bahcall \& Soneira (1983)'s
$6 \times 10^{-6} \hmpccc$ and with
Efstathiou et al. (1992)'s $7 \times 10^{-6} \hmpccc$.

Figure~\ref{fig_sky} shows the distributions of
clusters on the sky with measured redshifts in each catalog
used in our CF analysis.


\subsection{The APM Clusters} \label{apmdata}

Dalton et al. (1997) have published a catalog of APM clusters with
richness range ${\cal R} \ge 40$.
This is the catalog that Dalton et al. (1994a) have used to measure the APM
cluster CF, which is consistent with the results of Dalton et al. (1992) and Croft et al. (1997).
The survey region is shown by Maddox et al. (1990)\footnote{The exact APM survey
field definition can be found in the Stromlo-APM redshift catalog (\cite{lov96})
served by the Astronomical Data Center (http://adc.gsfc.nasa.gov/).}
and contains 957 clusters with $m_x \le 19.4$
($m_x$ is the apparent magnitude of $x$-th bright member galaxy
where $x = {\cal R} /2.1$; this is analogous to Abell cluster's $m_{10}$).
The number of clusters in the APM sample shows a rather strong
dependence on declination (Park \& Lee 1998).
In addition to the APM survey limits we have restricted our APM samples
to the declination range of $-25^\circ \ge \delta \ge -65^\circ$
to reduce the potential effects of this declination dependence.
But this change of declination boundaries turned out to make an insignificant
difference in the CF.

Since the APM galaxy catalog is complete over the magnitude range
$17.0 \le b_J \le 20.5$ (\cite{mad90})
and the cluster richness counting slice is
[$m_x - 0.5$, $m_x + 1.0$], the APM cluster
catalog is complete over magnitude range $17.5 \le m_x \le 19.4$.
This complete sample contains 927 clusters.
But, in this sample, the number of clusters with measured redshifts is only 350.
For clusters with ${\cal R} \ge 50$ and $17.5 \le m_x \le 19.2$, however,
the completeness becomes $83 \%$
and increases for higher richness limits.
We therefore make four magnitude-limited ($17.5 \le m_x \le 19.2$)
richness subsamples with ${\cal R} \ge 50$,
60 (completeness 93\%), 70 (96\%), and 80 (100\%)(see Table~\ref{table_op}).

To calculate the selection function of the APM clusters,
the bright magnitude limit $m_{x,low} = 17.5$
should be taken into account as well as the faint limit $m_{x,upp} = 19.2$.
The absolute magnitude $M_x$ corresponding to $m_x$
of each cluster is computed by equation~(3),
where $K(z) = 3z$ is adopted (\cite{dal97}).
Then the selection function of the APM survey is given by
\begin{equation}
S(d) = {3\over \Omega} \sum_{i; \atop {{D_{max,i} \ge d} \atop {D_{min,i} \le d}}}
 {1\over {D_{max,i}^{3} - D_{min,i}^{3}  }}
\end{equation}
where $D_{max,i}$ is the maximum distance to which an $i$-th cluster
can be observed by
the faint end of the survey magnitude $m_{x,upp} = 19.2$,
$D_{min,i}$ is the minimum distance
to which the cluster could be included in the sample of $m_{x,low} = 17.5$,
and $\Omega = 1.02$ sr is the solid angle of our survey region
of the APM sample.
Again, using the distribution of clusters smoothed in redshift space
as the selection function makes little change in our results.
The sample is limited to the distance interval
from $d_{in} = 50 \hmpc$ to $d_{out} = 500 \hmpc$.

The mean space densities of our APM ${\cal R} \ge 70$ and 80
subsamples agree well with those estimated by Croft et al. (1997).
But the density of our ${\cal R} \ge 50$ sample
is somewhat lower than $3.4 \times 10^{-5} \hmpccc$ of Croft et al. (1997)
because of lower completeness of clusters with published redshifts.
We adopt Croft et al. (1997)'s value
in the study of the richness dependence of correlation amplitudes
in \S~\ref{richdepend}.

\setcounter{figure}{2}
\begin{figure}[h]
\vspace{-1.3cm}
\centerline{\epsfysize=10cm\epsfbox{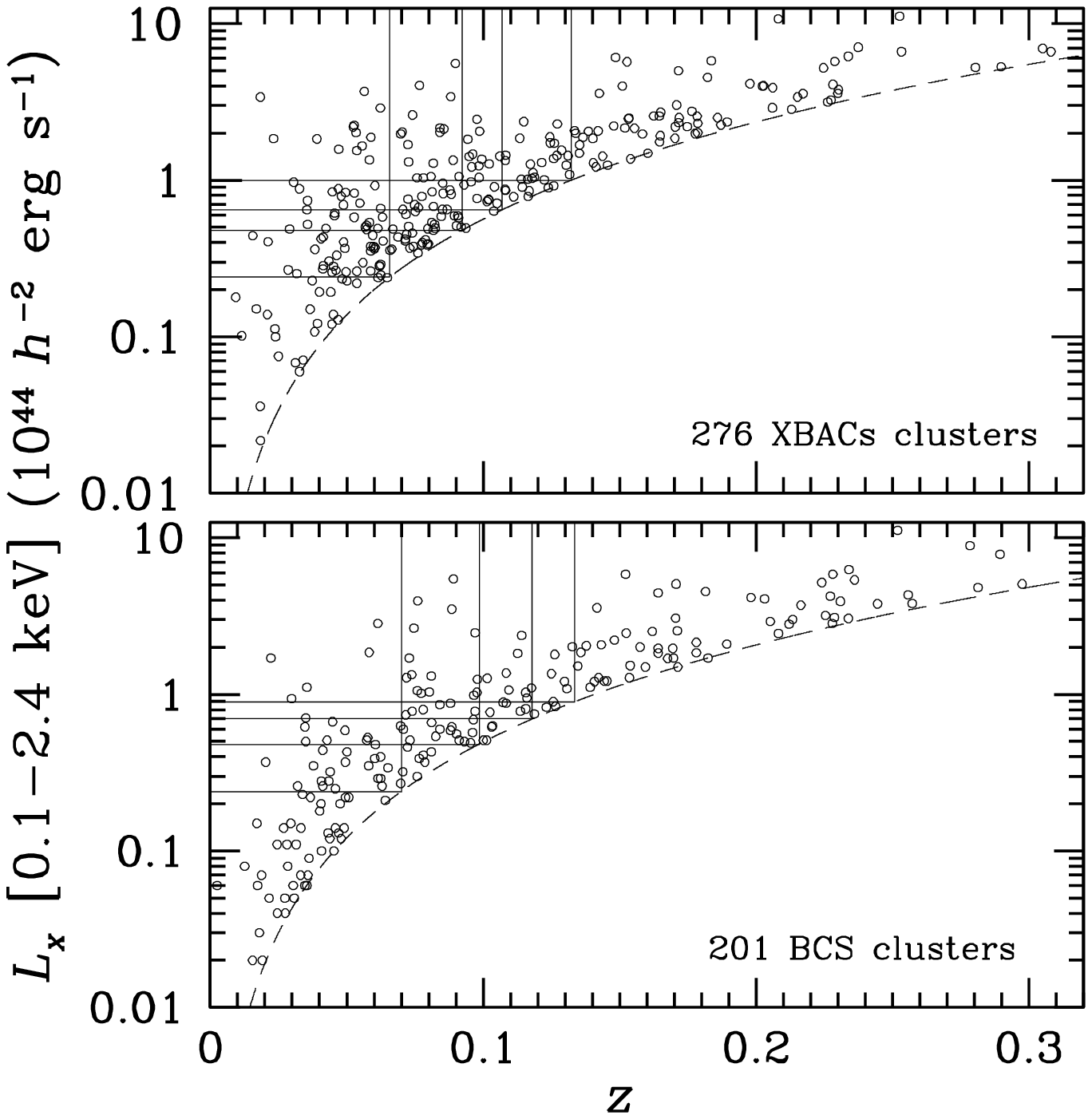}}
\vskip -1.5cm
\hskip 3.5cm {\begin{minipage}{8.4cm}\footnotesize
{\bf Fig.~2.}~---~X-ray luminosity $\lx$ versus redshift $z$ distribution of
276 XBACs clusters and 201 BCS clusters.
The dashed lines represent the $\lx$
limit corresponding to the survey flux limit;
$5.0 \fxcgs$ for the XBACs and
$4.4 \fxcgs$ for the BCS, respectively.
The solid lines show the
$\lx$ limits and distance limits of volume-limited subsamples.
\end{minipage}}
\vspace{0cm}
\end{figure}


\subsection{The XBACs} \label{xbacsdata}

The XBACs catalog (\cite{ebe96}) is a complete, all-sky,
X-ray flux-limited ($\fx \ge 5.0$ ${\rm \, \times 10^{-12} \;erg}$ ${\rm cm^{-2}\;s^{-1}}$
in the $0.1 \sim 2.4$ keV band) sample of 276
Abell clusters of galaxies compiled from the {\it ROSAT} All-Sky Survey.
We constrain our sample to $|b| \ge 30\arcdeg$ to reduce the effects
of the galactic obscuration. After the LMC and the SMC regions are removed,
the remaining survey area covers 6.27 sr of the sky.
The completeness of the XBACs catalog is about 80\% due to various sources
of incompleteness
which are very difficult to remove (\cite{ebe96}).
We have further removed the X-ray sources contaminated by point sources
or without measured redshifts from the XBACs catalog.
The six double clusters in the XBACs catalog are considered as
single systems which have undergone merging.
The resulting final sample includes 186 clusters
with the flux-limit of $\fx \ge 5.0 \fxcgs$ and with $|b| \ge
30\arcdeg$ in the redshift range $0<z<0.2$. This is 91\% of clusters
in the catalog with the same flux, galactic latitude, and redshift limits.

Assuming the Einstein-de Sitter universe we calculate
the X-ray luminosity $\lx$ of clusters from the observed flux $\fx$ and
redshift $z$. Four volume-limited subsamples with
$\lx \ge 0.24$, 0.48, 0.65 and 1.00 ${\rm \, \times 10^{44}\;} h^{-2}$
${\rm erg\;s^{-1}}$
and with the corresponding redshift limits of 0.066, 0.092, 0.107, and 0.132
are made (see Fig.~2). These samples contain 49, 67, 59,
and 51 clusters, respectively.
Our volume-limited XBACs subsamples contain many rich Abell clusters
with $RC > 1$.

\begin{deluxetable}{ccccc}
\tablecolumns{5}
\footnotesize
\tablecaption{
Volume-limited subsamples of the XBACs and the BCS clusters
\label{table_x}}
\tablewidth{0pt}
\tablehead{
\colhead{$L_X$} & \colhead{} &
\colhead{$n_c$} & \colhead{$r_0$} & \colhead{} \nl
\colhead{(${\rm 10^{44}\;} h^{-2} {\rm erg\;s^{-1}}$)} & \colhead{$N_c$} &
\colhead{($\!\hmpccc$)} & \colhead{($\!\hmpc$)} & \colhead{$\gamma$}
}
\startdata
\tablevspace{0.9mm}
\multicolumn{5}{c}{\small XBACs} \nl
\tableline
\tablevspace{1mm}
 $\lx \ge 0.24$ & 49 & $3.53 \times 10^{-6}$
& $25.7^{+3.7}_{-3.8}$ & $2.55^{+0.51}_{-0.43}$ \nl
\noalign{\smallskip}
 $\lx \ge 0.48$ & 67 & $1.84 \times 10^{-6}$
& $25.2^{+4.1}_{-4.3}$ & $2.48^{+0.55}_{-0.54}$ \nl
\noalign{\smallskip}
 $\lx \ge 0.65$ & 59 & $1.07 \times 10^{-6}$
& $30.3^{+8.2}_{-6.5}$ & $2.21^{+0.72}_{-1.27}$ \nl
\noalign{\smallskip}
 $\lx \ge 1.00$ & 51 & $0.53 \times 10^{-6}$
& $45.8^{+26.4}_{-12.5}$ & $1.84^{+0.75}_{-0.99}$ \nl
\tablevspace{0.7mm}
\tableline
\tablevspace{2mm}
\multicolumn{5}{c}{\small BCS} \nl
\tableline
\tablevspace{1mm}
 $\lx \ge 0.24$ & 33 & $3.02 \times 10^{-6}$
& $33.0^{+6.2}_{-5.9}$ & $1.82^{+0.49}_{-0.50}$ \nl
\noalign{\smallskip}
 $\lx \ge 0.48$ & 49 & $1.70 \times 10^{-6}$
& $24.9^{+5.4}_{-5.6}$ & $2.68^{+0.91}_{-0.77}$ \nl
\noalign{\smallskip}
 $\lx \ge 0.70$ & 40 & $0.84 \times 10^{-6}$
& $30.2^{+9.8}_{-8.9}$ & $2.22^{+0.94}_{-1.26}$ \nl
\noalign{\smallskip}
 $\lx \ge 0.90$ & 34 & $0.51 \times 10^{-6}$
& $36.3^{+21.8}_{-13.8}$ & $2.05^{+1.24}_{-1.76}$ \nl
\enddata
\end{deluxetable}


\subsection{The BCS} \label{bcsdata}

Recently, the {\it ROSAT} BCS catalog has been published (\cite{ebe98}).
Unlike the XBACs catalog which is based on the Abell cluster catalog,
the BCS clusters are selected purely by their X-ray properties.
The BCS contains 201 X-ray brightest clusters of galaxies
with measured redshifts in the northern hemisphere ($\delta \ge 0\arcdeg$)
and at $|b| \ge 20\arcdeg$, and is 90 \% complete at the X-ray flux limit of
$4.4 \fxcgs$ in the $0.1 \sim 2.4$ keV band.
Six sources, identified as three double clusters in the Abell catalog,
are considered as six independent X-ray clusters since the sample should
be based only on the X-ray information.
Adding these clusters and excluding three clusters with fluxes
lower than the flux limit, we have 201 X-ray clusters.

We have used the BCS to make four volume-limited subsamples with
$\lx \ge 0.24$, 0.48, 0.70, and 0.90 ${\rm \, \times 10^{44}\;}$
$h^{-2} {\rm erg\;s^{-1}}$ and with the redshift limits of
0.070, 0.099, 0.118, and 0.134 (see Fig.~2).
These subsamples contain 33, 49, 40, and 34 clusters, respectively.
Our high $\lx$-limited subsamples of the XBACs and the BCS clusters
contain very rare objects (see Table~\ref{table_x}).

\begin{figure*}[t]
\plotfiddle{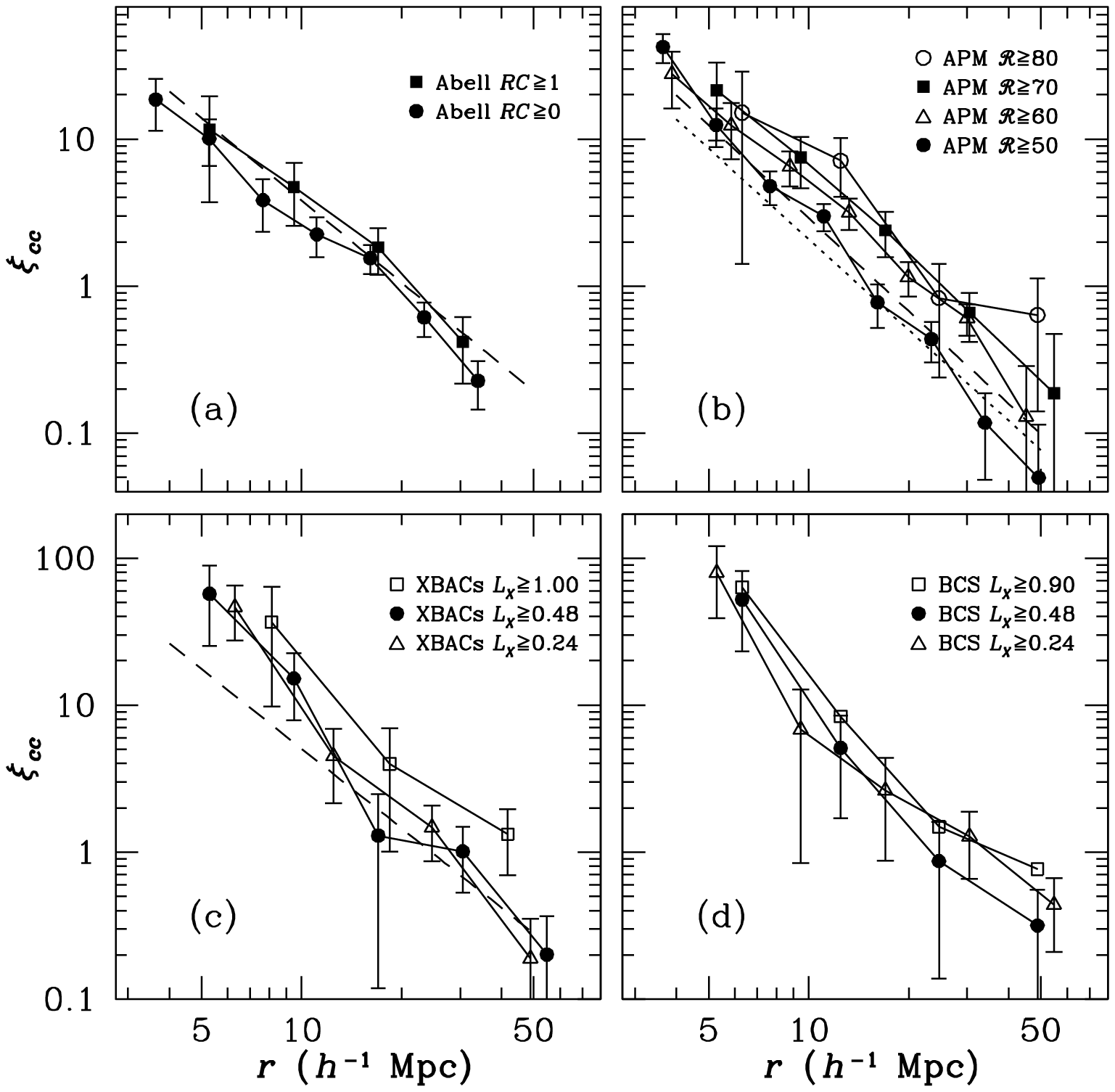}{10cm}{0}{67}{67}{-210}{-145}
\caption{\footnotesize
Correlation functions of clusters of galaxies.
({\it a}) Magnitude-limited samples of
the Abell clusters with $RC \ge 0$ and 1.
The dashed line is $\xi(r) = (r / 20.6\hmpc)^{-1.86}$ of
Postman et al. (1992) for $RC \ge 0$ Abell clusters.
({\it b}) Magnitude-limited samples of the APM clusters
with ${\cal R} \ge 50$, 60, 70, and 80.
The dotted line is $\xi(r) = (r / 14.3\hmpc)^{-2.05}$ of
${\cal R} \ge 50$ sample of Croft et al. (1997)
and the dashed line is $\xi(r) = (r / 16.6\hmpc)^{-2.10}$ of
their ${\cal R} \ge 70$ sample.
({\it c}) Volume-limited samples of
the XBACs clusters with $\lx \ge 0.24$, 0.48,
and $1.00 \lxcgs$.
The dashed line is $\xi(r) = (r / 24.6\hmpc)^{-1.80}$ of
Abadi et al. (1998) for clusters with $\lx \ge 0.54 \lxcgs$.
({\it d}) Volume-limited samples of the BCS clusters with $\lx \ge 0.24$, 0.48,
and $0.90 \lxcgs$.
\label{fig_xi}}
\end{figure*}


\section{CORRELATION FUNCTIONS} \label{xicc}

The correlation function is an estimate of excess clustering over
the random Poisson distribution.
We estimate the CFs using the conventional estimator DD/DR.
We have also used the Hamilton (1993)'s estimator
\begin{equation}
\xi_{cc}(r) = {\rm{{DD \cdot RR} \over {DR^2}}}
{{4N_cN_r} \over {(N_c - 1)(N_r - 1)}} - 1,
\end{equation}
which is less affected by uncertainties in the selection
function for $\xi_{cc} < 1$ compared to the DD/DR-method.
We find that the differences between two estimators are
insignificant in our analysis.
Here  DD is the number of pairs at separation $r$
in the sample with $N_c$ clusters, RR is the number of pairs in a
random sample with $N_r$ random points, occupying the same survey volume and
with the same selection function as the real cluster sample, and DR is the
cluster-random pair count.
We compute the uncertainties of $\xi_{cc}$ using the formula
$\delta \xi_{cc} = (1 + \xi_{cc}) / \sqrt{{\rm DD}}$.
Table~\ref{table_cf} lists the CF data we have calculated.


\subsection{Optically Selected Clusters} \label{xiccop}

Least-square fits of a power law equation~(1) to the best estimates
of CF are presented in Table~\ref{table_op}
for the richness subsamples of the Abell and the APM clusters.
We determine the 68.3\% confidence limits of each parameter individually
assuming that the distribution for the uncertainties of $\xi_{cc}$ is
Gaussian and different $\xi_{cc}$ bins are independent of each other,
which is a popular, but strictly speaking, imprecise assumption.

\begin{deluxetable}{cc@{}ccc@{}ccc@{}ccc@{}ccc@{}ccc}
\tablecolumns{17}
\scriptsize
\tablecaption{
Two-point correlation functions of clusters of galaxies
\label{table_cf}}
\tablewidth{0pt}
\tablehead{
\tablevspace{1.5mm}
\multicolumn{5}{c}{\small Abell} & & \multicolumn{11}{c}{\small APM} \nl
\cline{1-5} \cline{7-17} \nl
\multicolumn{2}{c}{\scriptsize $RC \ge 0$} & &
\multicolumn{2}{c}{\scriptsize $RC \ge 1$} & &
\multicolumn{2}{c}{\scriptsize ${\cal R} \ge 50$} & &
\multicolumn{2}{c}{\scriptsize ${\cal R} \ge 60$} & &
\multicolumn{2}{c}{\scriptsize ${\cal R} \ge 70$} & &
\multicolumn{2}{c}{\scriptsize ${\cal R} \ge 80$} \nl
\tablevspace{1mm}
\cline{1-2} \cline{4-5} \cline{7-8} \cline{10-11} \cline{13-14} \cline{16-17} \nl
\colhead{\scriptsize $r$\tablenotemark{a}} & \colhead{\scriptsize $\xi_{cc}(r)$} & &
\colhead{\scriptsize $r$} & \colhead{\scriptsize $\xi_{cc}(r)$} & &
\colhead{\scriptsize $r$} & \colhead{\scriptsize $\xi_{cc}(r)$} & &
\colhead{\scriptsize $r$} & \colhead{\scriptsize $\xi_{cc}(r)$} & &
\colhead{\scriptsize $r$} & \colhead{\scriptsize $\xi_{cc}(r)$} & &
\colhead{\scriptsize $r$} & \colhead{\scriptsize $\xi_{cc}(r)$}
}
\startdata
\phn3.6 & $19 \pm  7\phn$ & &
\phn5.3 & $12 \pm  8\phn$ & &
\phn3.6 & $42 \pm 10$ & &
\phn3.9 & $28 \pm 12$ & &
\phn5.3 & $21 \pm 12$ & &
\phn6.3 & $15 \pm 14$ \nl
\phn5.3 & $10 \pm 3\phn$ & &
\phn9.5 & $4.7 \pm 2.2$ & &
\phn5.3 & $12 \pm 4\phn$ & &
\phn5.8 & $12 \pm 5\phn$ & &
\phn9.5 & $7.5 \pm 2.9$ & &
   12.5 & $7.1 \pm 3.1$ \nl
\phn7.7 & $3.8 \pm 1.5$ & &
   17.0 & $1.8 \pm 0.6$ & &
\phn7.7 & $4.8 \pm 1.2$ & &
\phn8.8 & $6.5 \pm 1.7$ & &
   17.0 & $2.4 \pm 0.8$ & &
   24.7 & $0.83 \pm 0.59$ \nl
   11.1 & $2.3 \pm 0.7$ & &
   30.5 & $0.42 \pm 0.20$ & &
   11.1 & $3.0 \pm 0.6$ & &
   13.2 & $3.2 \pm 0.8$ & &
   30.5 & $0.66 \pm 0.24$ & &
   48.8 & $0.64 \pm 0.50$ \nl
   16.1 & $1.5 \pm 0.3$ & &
\nodata & \nodata & &
   16.1 & $0.77 \pm 0.26$ & &
   19.9 & $1.2 \pm 0.3$ & &
   54.8 & $0.19 \pm 0.28$ & &
\nodata & \nodata \nl
   23.4 & $0.61 \pm 0.16$ & &
\nodata & \nodata & &
   23.4 & $0.44 \pm 0.13$ & &
   30.0 & $0.61 \pm 0.15$ & &
\nodata & \nodata & &
\nodata & \nodata \nl
   33.9 & $0.23 \pm 0.08$ & &
\nodata & \nodata & &
   33.9 & $0.12 \pm 0.07$ & &
   45.2 & $0.13 \pm 0.16$ & &
\nodata & \nodata & &
\nodata & \nodata \nl
\nodata & \nodata & &
\nodata & \nodata & &
   49.2 & $0.050 \pm 0.064$ & &
\nodata & \nodata & &
\nodata & \nodata & &
\nodata & \nodata \nl
\tablevspace{0.5mm}
\tableline
\nl \nl
\tableline
\tableline \nl
\multicolumn{8}{c}{\small XBACs} & & \multicolumn{8}{c}{\small BCS} \nl
\cline{1-8} \cline{10-17}
\tablevspace{2.2mm}
\multicolumn{2}{c}{\scriptsize $\lx \ge 0.24$\tablenotemark{b}} & &
\multicolumn{2}{c}{\scriptsize $\lx \ge 0.48$} & &
\multicolumn{2}{c}{\scriptsize $\lx \ge 1.00$} & &
\multicolumn{2}{c}{\scriptsize $\lx \ge 0.24$} & &
\multicolumn{2}{c}{\scriptsize $\lx \ge 0.48$} & &
\multicolumn{2}{c}{\scriptsize $\lx \ge 0.90$} \nl
\tablevspace{1mm}
\cline{1-2} \cline{4-5} \cline{7-8} \cline{10-11} \cline{13-14} \cline{16-17} \nl
\colhead{\scriptsize $r$} & \colhead{\scriptsize $\xi_{cc}(r)$} & &
\colhead{\scriptsize $r$} & \colhead{\scriptsize $\xi_{cc}(r)$} & &
\colhead{\scriptsize $r$} & \colhead{\scriptsize $\xi_{cc}(r)$} & &
\colhead{\scriptsize $r$} & \colhead{\scriptsize $\xi_{cc}(r)$} & &
\colhead{\scriptsize $r$} & \colhead{\scriptsize $\xi_{cc}(r)$} & &
\colhead{\scriptsize $r$} & \colhead{\scriptsize $\xi_{cc}(r)$} \nl
\tablevspace{0.6mm}
\tableline
\tablevspace{1.3mm}
\phn6.3 & $46 \pm 19$ & &
\phn5.3 & $57 \pm 32$ & &
\phn8.1 & $37 \pm 27$ & &
\phn5.3 & $79 \pm 41$ & &
\phn6.3 & $52 \pm 29$ & &
\phn6.3 & $63 \pm 70$ \nl
   12.5 & $4.5 \pm 2.4$ & &
\phn9.5 & $15 \pm 7\phn$ & &
   18.4 & $4.0 \pm 3.0$ & &
\phn9.5 & $6.8 \pm 6.0$ & &
   12.5 & $5.1 \pm 3.4$ & &
   12.5 & $8.4 \pm 9.3$ \nl
   24.7 & $1.5 \pm 0.6$ & &
   17.0 & $1.3 \pm 1.2$ & &
   41.8 & $1.3 \pm 0.6$ & &
   17.0 & $2.6 \pm 1.8$ & &
   24.7 & $0.87 \pm 0.73$ & &
   24.7 & $1.5 \pm 1.8$ \nl
   48.8 & $0.19 \pm 0.16$ & &
   30.5 & $1.0 \pm 0.5$ & &
\nodata & \nodata & &
   30.5 & $1.3 \pm 0.6$ & &
   48.8 & $0.32 \pm 0.24$ & &
   48.8 & $0.77 \pm 0.59$ \nl
\nodata & \nodata & &
   54.8 & $0.20 \pm 0.17$ & &
\nodata & \nodata & &
   54.8 & $0.44 \pm 0.23$ & &
\nodata & \nodata & &
\nodata & \nodata \nl
\enddata
\tablenotetext{a}{\footnotesize $\,$Separation in units of $h^{-1}{\rm Mpc}$.}
\tablenotetext{b}{\footnotesize $\,$X-ray luminosity in units of
${\rm 10^{44}\;} h^{-2} {\rm erg\;s^{-1}}$.}
\end{deluxetable}

Figure~\ref{fig_xi}a shows the CFs of the Abell clusters with
$RC \ge 0$ (circle) and $RC \ge 1$ (square).
The dashed line is Postman et al. (1992)'s power law
fit with $r_0 = 20.6 \hmpc$ and
$\gamma = 1.86$ estimated for the Abell $RC \ge 0$ sample.
The CFs in Figure~\ref{fig_xi}b are those of
the APM clusters with ${\cal R} \ge 50$ (filled circle),
${\cal R} \ge 60$ (triangle),
${\cal R} \ge 70$ (square), and ${\cal R} \ge 80$ (open circle).
The dotted line is a power law with $r_0 = 14.3 \hmpc$ and $\gamma = 2.05$
(${\cal R} \ge 50$)
and the dashed line is with $r_0 = 16.6 \hmpc$ and $\gamma = 2.10$
(${\cal R} \ge 70$)
both determined by Croft et al. (1997).
It can be noted
that the amplitude of CF steadily increases as the richness increases.

Dalton et al. (1992) have compared the richness of the ACO clusters with that of
the APM clusters (hereafter APM92) and found that the mean difference
in richness is $\langle R_{\rm ACO} - {\cal R}_{\rm APM92} \rangle = 10.1$.
And Dalton et al. (1994a) have compared a new version of the APM cluster catalog
(which is the catalog we use) with their old version catalog, and found
a relation ${\cal R}_{\rm APM} = 1.2 {\cal R}_{\rm APM92} + 24.8$
with a scatter of 16 in ${\cal R}_{\rm APM}$. We then
get a rough relation ${\cal R}_{\rm APM} = 1.2 R_{\rm ACO} +12.7$.
Therefore, the counts of the Abell clusters with
$R = 30$ ($RC=0$) and 50 ($RC=1$)
correspond to the richnesses of the APM clusters with roughly
${\cal R} = 49$ and 73, respectively.
In Table~\ref{table_op} we find that the correlation lengths of the APM clusters
are generally similar to those of the Abell clusters
at similar levels of richness.
Contrary to Croft et al. (1997), the APM subsamples with higher richness
limits show clustering as strong as or rather stronger than
the Abell cluster samples.
It should be noted that there are
large uncertainties both in the richness relation and in the richness itself
(the uncertainties of $R_{\rm ACO}$ and  ${\cal R}_{\rm APM}$
is about 18 and 5, respectively; \cite{dal97}), and that the Abell $RC\ge 0$
sample is an incomplete sample with fewer poorer clusters.

Park \& Lee (1998) have calculated the CFs
of the APM clusters using various CF calculation methods and sample
corrections. They have used different selection functions, varied
sample boundaries both in angle and depth, and applied galactic latitude
or declination dependence corrections. Even with these variations
they have consistently obtained CFs with amplitude higher than
that reported by  Dalton et al. (1994a) or by Croft et al. (1997),
and could not find the sources for discrepancy.
Dalton (1998) has pointed out that our redshift
sample contains the clusters with redshifts drawn from the Las Campanas
Redshift survey or from the Edinburgh-Durham-Milano cluster redshift survey
which cover only parts of the APM Galaxy Survey region, and that these
clusters could produce an inhomogeneity in the distribution of clusters
on the sky.
Our statistical samples, however, contain very few of those clusters,
and we have found no effect on the CF by them.
For example, our magnitude-limited sample with ${\cal R} \ge 70$
does not contain any of the clusters in question.


\subsection{X-ray Clusters} \label{xiccx}

Figure~\ref{fig_xi}c shows the CFs of the XBACs samples with
$\lx \ge 0.24$ (triangle), 0.48 (circle), and 1.00
${\rm \, \times 10^{44}\;} h^{-2}$ ${\rm erg\;s^{-1}}$ (square).
The dashed line represents Abadi et al. (1998)'s fit
$\xi(r) = (r/24.6\hmpc)^{-1.80}$ for the XBACs clusters
with $\lx \ge 0.54 \lxcgs$.
The correlation lengths reported by Abadi et al. (1998) agree well with
our results, except for the highest $\lx$ subsample with $\lx \ge 1.00 \lxcgs$.
It should be considered, however, that the uncertainties in the
fitting parameters are very large both for their and our
highest $\lx$ subsamples.
The CFs of the BCS clusters are shown in Figure~\ref{fig_xi}d.

In Table~\ref{table_x} we list the parameters of CFs of
the XBACs and the BCS clusters.
Our X-ray cluster subsamples have very low number density
and consist of the rarest clusters which might be very massive.
As expected, the correlation length is very large and
tends to increase as the luminosity limit increases.
Borgani, Plionis, \& Kolokotronis (1999) calculated the CFs from
the XBACs and their resulting $r_0$ from the sample with the same flux-limit as ours,
$\fx \ge 5.0 \fxcgs$, is $26.0 \hmpc$,
which agrees excellently with our result of the lowest $\lx$ subsample; $r_0 = 25.7 \hmpc$.

\begin{figure*}[t]
\plotfiddle{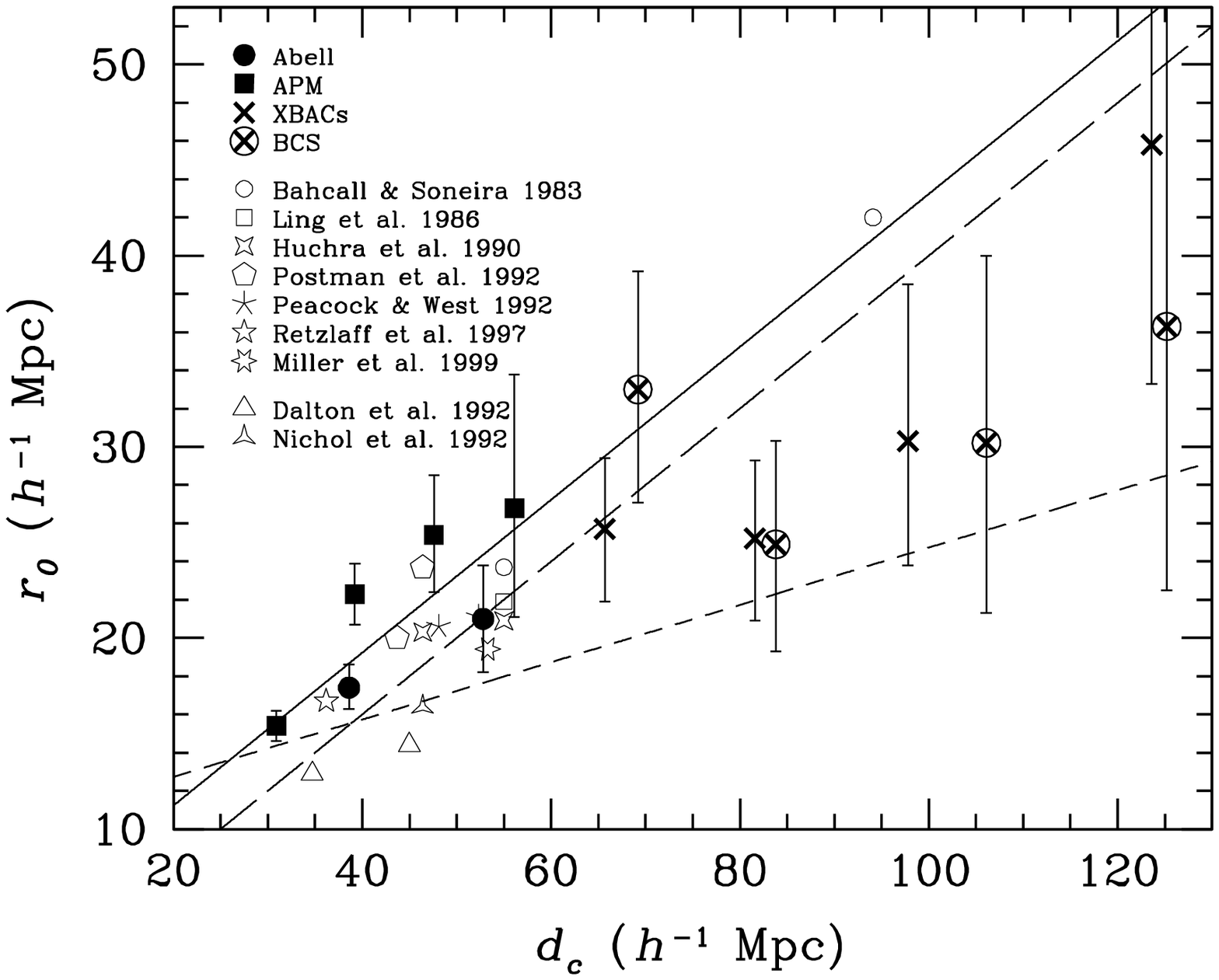}{11.0cm}{0}{80}{80}{-230}{-195}
\caption{\footnotesize
Dependence of the correlation length $r_0$
on the richness or the mean intercluster separation $d_c$.
The solid line is $r_0 = 0.40 d_c + 3.2$, which is
a best fit to our results for the Abell and the APM samples.
The long dashed line is the relation reported by Bahcall \& West (1992),
and the short dashed line is by Croft et al. (1997).
\label{fig_rich}}
\end{figure*}


\section{RICHNESS DEPENDENCE} \label{richdepend}

We compare the amplitudes of CFs of the Abell and
the APM clusters having the same level of richness.
Richness level of a cluster sample is often represented by
the mean separation $d_{c} = n_{c}^{-1/3}$ rather than richness itself
(\cite{sza85}; \cite{bah88}; \cite{baw92}; \cite{cro97})
because the definition on richness is different for different catalogs.

Our results on the richness dependence of the cluster CF
are shown in Figure~\ref{fig_rich}. Filled circles are the relations
measured from the Abell sample, and filled squares are from the APM sample.
We find that the clustering strength of the APM clusters
is consistent with that of the Abell clusters, contrary to what
Croft et al. (1997) have claimed.
On the other hand, our results agree well with the previous studies
on the Abell clusters (see Table~1 of \cite{nic94}
and other references presented in Fig.~\ref{fig_rich}),
but disagree with those of the APM team and Nichol et al. (1992).
The solid line in Figure~\ref{fig_rich} is a linear fit
to the $r_0$ versus $d_c$ distribution found from our optical cluster samples.
The relation we found is $r_0 = 0.40 d_c + 3.2$.
The long and short dashed lines are the relations of Bahcall \& West (1992)
and Croft et al. (1997), respectively.
Our relation is very similar to Bahcall \& West (1992)'s law
except for a constant offset.
However, our result strongly disagrees with the slope found by
Croft et al. (1997).

The $r_0$ versus $d_c$ relations of the XBACs and
the BCS clusters are consistent with each other,
but are much lower than the relation for optical clusters.
The slope of the relation is uncertain, but tends to increase
as $d_c > 80 \hmpc$.
The correlation lengths calculated by Borgani et al. (1999) from the XBACs
and by Collins et al. (2000) from
the ROSAT-ESO Flux-Limited X-ray (REFLEX) galaxy cluster survey (\cite{boh01})
are consistent with our results
and their distributions show rather weak dependence with $\lx$ like ours,
or look like even randomly distributed.
As discussed in the following section,
we think the X-ray luminosity of rich
clusters is not a very good measure of richness or mass.
It should be also noted that the uncertainties in $r_0$ are rather large.


\section{DISCUSSION} \label{discuss}

We have measured the cluster-cluster correlation function
by using four catalogs of optical and X-ray clusters.
We have applied the same analysis method and the same method of
characterizing the resulting CFs to all observational samples.
Our CF for the Abell cluster sample is consistent with
the previous works by Batuski \& Burns (1985), Ling, Frenk, \& Barrow (1986),
Retzlaff et al. (1998), and Miller et al. (1999).
The correlation lengths reported by Postman et al. (1992) and
Bahcall \& Soneira (1983) are slightly larger than ours.
Those estimated by Huchra et al. (1990) and
Peacock \& West (1990) are very close to ours for the Abell clusters
with $RC \ge 1$, but are somewhat larger than ours in the case of $RC \ge 0$.
On the other hand, our CF for the APM cluster sample has the correlation length
significantly larger than those reported by Dalton et al. (1992, 1994a) and
Croft et al. (1997), in particular for richer subsamples.
We have not been able to find the reason for this discrepancy (\cite{par98}).

Sutherland (1988) and Efstathiou et al. (1992) have suggested that
projection effects of the Abell cluster catalog caused an artificial
radial correlation, which could lead to an overestimation of CF
relative to the APM sample.
They have attempted to correct the Abell sample for the projection effects,
and reported that
the correlation length was reduced to $r_0 \simeq 14 \hmpc$.
However, several other studies have pointed out that the uncertainties
inherent in this correction are very large, and that the correction
may have been overdone (\cite{nic92}).
Miller et al. (1999) found that the excess radial correlation
is not artificial but due to two elongated superclusters, Ursa Majoris and Corona Borealis.
Postman et al. (1992) have estimated
the magnitude of the projection contamination and have found that
only 17 clusters (5\%) in their 351 Abell clusters may be contaminated.
They also have measured the effects of these contaminated clusters
on the CF and have found that the amplitude and slope of
the CF do not change significantly.
Our study indicates that the APM team's CFs of the APM clusters
are underestimated for richer subsamples, and the CF of the Abell sample
is consistent with that of the APM sample at similar richness.

We have detected a rather strong dependence of the clustering amplitude
on the richness of clusters.
The correlation length $r_0$ versus mean separation $d_c$
relation we have found from the Abell and the APM samples,
$r_0 = 0.40 d_c + 3.2$, has a slope equal to that found by
Bahcall \& West (1992), but has a finite correlation length at zero
separation. We think this is more reasonable because the correlation
length should be finite at the continuum limit, namely for the matter field.

By using the X-ray selected clusters one can greatly
overcome the projection effects which might have affected the optical cluster
samples selected from the distribution of galaxies projected on the sky.
Although the XBACs clusters are not genuine X-ray selected clusters
and have incompleteness due to the missing poor ACO clusters,
they are indeed three-dimensionally bound systems and the sampling is
nearly free from false detection or overestimation of richness.
The BCS clusters are selected purely from their X-ray properties,
and also have negligible projection effects.
However, since the number of X-ray clusters is not yet large enough to
make a sufficiently large volume-limited sample,
the CFs measured from  the XBACs and
the BCS clusters have large uncertainties compared
to those of optical clusters. The clustering strength of these X-ray clusters
is in general lower than what is expected from the $r_0$ versus $d_c$
relation of optical clusters.
This is probably due to the fact that the X-ray luminosity of X-ray clusters
is not an excellent measure of the cluster richness or mass.
The X-ray luminosity of a cluster depends on
the environment and activity as well as its dynamical mass.
In fact, Figure 4 of David, Forman \&
Jones (1999) shows that the clusters very bright
in X-ray belong to various richness classes.
Gilbank, Bower, \& Castander (2001) also found that
X-ray selection misses some significant rich clusters
while optical selection can detect all X-ray clusters.
However, there is a trend of the clustering strength
of the X-ray clusters increasing as the mean separation increases.
Despite the large uncertainties,
the correlation length of the X-ray clusters
is higher in amplitude compared to Croft et al. (1997)'s expectation.

We conclude that the Abell and the APM clusters
do have CFs statistically consistent with each other and show
a similar richness dependence relation of the clustering strength.
It should be again emphasized that our results have been obtained by
applying the  same method of analyzing the
observational data and calculating the CF.


\acknowledgments{
We thank Dr. Ebeling for supplying us with the BCS catalog at the
earliest time. We also thank Drs. Michael S. Vogeley, Neta A. Bahcall,
Gavin B. Dalton, and R. A. C. Croft for valuable comments.
This work was supported by the Basic Science Research Institute
Program, Ministry of Education 1995 (BSRI-98-5408).
}



\end{document}